\documentclass[twocolumn]{aastex631}

\newcommand\realfast{\emph{realfast}}
\newcommand{\dmunits}{pc~cm$^{-3}$}
\newcommand{\rmunits}{rad~m$^{-2}$}

\usepackage{multirow}
\usepackage{amsmath}
\usepackage{longtable}
\usepackage{graphicx}

\usepackage{rotating} 
\begin{document}

\title{An unidentified \textit{Fermi} source emitting radio bursts in the Galactic bulge}
\shorttitle{Unidentified \realfast\ source}
\shortauthors{Anna-Thomas et al.}

\author[0000-0001-8057-0633]{Reshma Anna-Thomas}
\affiliation{West Virginia University, Department of Physics and Astronomy, P. O. Box 6315, Morgantown, WV, USA}
\affiliation{Center for Gravitational Waves and Cosmology, West Virginia University, Chestnut Ridge Research Building, Morgantown, WV, USA}
\correspondingauthor{Reshma Anna-Thomas}
\email{rat0022@mix.wvu.edu}

\author[0000-0003-4052-7838]{Sarah Burke-Spolaor}
\affiliation{West Virginia University, Department of Physics and Astronomy, P. O. Box 6315, Morgantown, WV, USA}
\affiliation{Center for Gravitational Waves and Cosmology, West Virginia University, Chestnut Ridge Research Building, Morgantown, WV, USA}
\affiliation{Sloan Fellow}

\author[0000-0002-4119-9963]{Casey J. Law}
\affiliation{Cahill Center for Astronomy and Astrophysics, MC 249-17 California Institute of Technology, Pasadena CA 91125, USA}
\affiliation{Owens Valley Radio Observatory, California Institute of Technology, Big Pine CA 93513, USA}

\author[0000-0001-6672-128X]{F.K.\ Schinzel}
\altaffiliation{An Adjunct Professor at the University of New Mexico.}
\affiliation{National Radio Astronomy Observatory, P.O. Box O, Socorro, NM 87801, USA}

\author[0000-0002-2059-0525]{Kshitij Aggarwal}
\affiliation{West Virginia University, Department of Physics and Astronomy, P. O. Box 6315, Morgantown, WV, USA}
\affiliation{Center for Gravitational Waves and Cosmology, West Virginia University, Chestnut Ridge Research Building, Morgantown, WV, USA}

\author[0000-0003-4056-9982]{Geoffrey C.\ Bower}
\affiliation{Academia Sinica Institute of Astronomy and Astrophysics, 645 N. A'ohoku Place, Hilo, HI 96720, USA}

\author[0000-0002-7587-6352]{Liam Connor}
\affiliation{Cahill Center for Astronomy and Astrophysics, MC 249-17 California Institute of Technology, Pasadena CA 91125, USA}
\affiliation{Owens Valley Radio Observatory, California Institute of Technology, Big Pine CA 93513, USA}

\author[0000-0002-6664-965X]{Paul B.\ Demorest}
\affiliation{National Radio Astronomy Observatory, P.O. Box O, Socorro, NM 87801, USA}

\begin{abstract}

We report on the detection of radio bursts from the Galactic bulge using the real-time transient detection and localization system, \realfast. The pulses were detected commensally on the Karl G. Jansky Very Large Array during a survey of unidentified \textit{Fermi} $\gamma$-ray sources. The bursts were localized to subarcsecond precision using \realfast\ fast-sampled imaging. Follow-up observations with the Green Bank Telescope detected additional bursts from the same source. The bursts do not exhibit periodicity in a search up to periods of 480 s, assuming a duty cycle of $<$ 20\%. The pulses are nearly 100\% linearly polarized, show circular polarization up to 12\%, and exhibit variable scattering on timescales of months. The arcsecond-level \realfast\ localization links the source confidently with the \textit{Fermi} $\gamma$-ray source and places it nearby (though not coincident with) an XMM-Newton X-ray source. Based on the source's overall properties, we discuss various options for the nature of this object and propose that it could be a young pulsar, a magnetar, or a binary pulsar system.
\end{abstract}

\section{Introduction} \label{sec:intro}
Astrophysical transients are events whose duration ranges between milliseconds to years. They inform us about the most variable and energetic events in the Universe.
Radio transients are typically described as ``fast" if they last for $<$ 1 s \citep{LawNgVLA}. Some of the fast radio transients, including pulsars \citep{Hewish1968}, rotating radio transients (RRATs; \citealt{Mclaughlin2006}), magnetar bursts, and fast radio bursts (FRBs) \citep{Lorimer2007}, are millisecond-duration emissions that are energetic enough to invoke coherent emission mechanisms to describe their high brightness temperature ($\rm T_{B}\ge 10^{30}$ K) and luminosities. Most of these are produced by highly magnetized neutron stars, even though the origin of FRBs is highly debated. Pulsars, RRATs, and magnetars have a rotational period, while FRBs do not. FRBs are also many orders of magnitude brighter and more energetic than Galactic pulsars, magnetars, or RRATs, with the exception of
FRB20200120E which has emitted bursts less luminous than FRB-like event FRB~200428 from the
Galactic magnetar SGR~1935+2154. Nevertheless, these transients are unique probes in understanding the baryonic content of the Milky Way interstellar medium (ISM) and the intergalactic medium (IGM), in the case of FRBs \citep{Macquart2020}, along their line-of-sight (LOS). Regular timing of millisecond pulsars (MSPs) has been used to detect stochastic gravitational wave background in the Universe \citep{HellingsDowns,nanograv}. Multiple ways have been proposed to detect these radio transients, including all-sky blind searches. One such method is to look for radio counterparts in the known high-energy sources, as some of these sources, like pulsars and magnetars, are established high-energy emitters.

\textit{Fermi} is a space mission that studies the cosmos in the energy range 10 keV--300 GeV. The imaging telescope on \textit{Fermi} called the Large Area Telescope (LAT) has a wide field of view ($>$8000\,cm$^{-2}$) and an angular resolution of $<3.5\deg$. \textit{Fermi}-LAT (4FGL-DR4) reported about 7195 point sources in their latest data release \citep{Fermi4FGL-DR4}. About 2427 (one-third) of these \textit{Fermi} sources don't have counterparts in any other electromagnetic regime. The origin and nature of these unassociated sources remain a mystery. A large fraction of these are eventually expected to be associated with the largest associated source classes, active galactic nuclei or pulsars. Recently, there appears a third class of so-called soft Galactic unassociated sources that are found in high-density regions of the Galactic plane. For a more detailed discussion of the unassociated gamma-ray source population see (\cite{Fermi4FGL-DR3}). \textit{Fermi}-LAT has detected at least 297 $\gamma$-ray pulsars, 70\% of which 
are radio-loud, and MSPs and young pulsars dominate this sample \citep{Smith2023}. Therefore, searching for new pulsars in the \textit{Fermi} sources remains a reasonable case. 

 The potential for precise localization capabilities make radio interferometers well-suited for searching for radio transients. Radio imaging-based pulsar searches have led to the discovery of the first isolated MSPs \citep{Erickson1980} and the first globular cluster pulsar \citep{Hamilton1985}. This technique is based on the assumption that pulsars are compact and steep spectrum radio sources. Such searches have successfully led to the discovery of many MSPs and normal pulsars \citep{Frail2016,Bhakta2017}. 

\realfast\ is a real-time commensal transient search system deployed at the Karl G. Jansky Very Large Array (JVLA) \citep{Law2015,realfast}.  It makes real-time interferometric images on visibility data sampled at 10 ms duration. This has the advantage of simultaneously detecting and localizing the transient to sub-arcsecond precision. \realfast\ has so far played a crucial role in the localization and host galaxy determination of many FRBs, including the first repeating FRB 20121102A \citep{Chatterjee2017}, FRB 20180916B \citep{Aggarwal2020}, FRB 20180301A \citep{Bhandari2022}, FRB 20201124A \citep{Ravi2022}, FRB 20200120E \citep{Kirtsen2022} and FRB 20190520B \citep{Niu2022}. It has also discovered a non-repeating FRB 20190614D in commensal during a VLA observation \citep{Law2020}. In this paper, we report on the discovery and localization of a Galactic radio bursting source, associated with a \textit{Fermi} $\gamma$-ray source, discovered by \realfast. 

The remaining of the paper has been organized into multiple sections. The observations, data reduction, and detection of the radio source are described in Section \ref{sec:data}. Section \ref{sec:periodicity} outlines the periodicity searches we did on the data, Section \ref{sec:properties} describes the properties of the bursts, Section \ref{sec:discussion} discusses the possible nature of the source and Section \ref{sec:conclusion} summarizes the results.
\section{Data and Burst Detection} \label{sec:data}
\subsection{Discovery and \textit{Fermi} Coincidence}
The source in this paper, called J1818--1531 hereafter, was detected on 2019 September 2 (MJD 58728) as a part of VLA program SC1046, which imaged the regions near unidentified \textit{Fermi} $\gamma$-ray sources in the inner Galaxy. The VLA was in the A configuration with 27 antennas and 351 baselines. The observation was done at VLA L-Band at a center frequency of 1.4~GHz. The total bandwidth of 1024 MHz was divided into 16 spectral windows, each with 64 channels having 1 MHz channel bandwidth. The main goal of the project was to identify steep-spectrum radio counterparts associated with the \textit{Fermi} sources. 
While the telescope was pointed at RA=18:18:37.91 and DEC=-15:33:41.39 for one of the SC1046 targets, the \realfast\ system detected five pulses at 1.4~GHz. The measured properties of these pulses are discussed below and presented in Table~\ref{tab:realfastbursts}, and had an average dispersion measure (DM) of around 1016\,\dmunits. The pulse position was coincident with \textit{Fermi} 95\% $\gamma$-ray error ellipses of 3FGL J1818.7-1528 using 4 years of \textit{Fermi} data \citep{Fermi3FGL}, and 4FGL J1818.6-1533 using eight years of \textit{Fermi} data \citep{Fermi4FGL}. The $\gamma$-ray source 4FGL\,J1818.6-1533 has two analysis flags (8196), making it part of the group of soft Galactic unassociated sources with higher flux uncertainty due to changes with older model or analysis. The DM-derived distance is 10.9 kpc using the NE2001 electron density model \citep{ne2001}, and 5.8 kpc using the YMW16 model \citep{ymw16}. This puts the source at a distance of 3.7 (3.03) kpc away from the Galactic Center for NE2001 (YMW16) distances.
\subsubsection{Burst searching and localization}
The \realfast\ search system was running alongside the SC1046 VLA observations. The visibilities are correlated commensally and sampled at a 10-ms resolution. These visibilities are then given to the graphics processing unit cluster on which the search pipeline \texttt{rfpipe} applies calibration, dedisperses, and forms images at many trial widths and DMs. The 8$\sigma$ fluence threshold limit of a 10-ms image is 0.29 Jy ms for the L-band. If the image S/N is greater than the threshold, the fast-sampled visibilities, 2 to 5 s that include the candidate, are recorded. The frequency-time data averaged over visibilities for each candidate is processed and classified using the machine learning classifier \textsc{Fetch} \citep{Agarwal2020}. The \realfast\ image of the candidates is convolved with the point spread function and is calibrated in real-time. The real-time images are made with several assumptions, including coarse DM grid, non-optimal image size, simpler calibration, etc. To resolve this, we used the raw visibilities dedispersed at the real-time detected DM to make burst images with CASA. The visibilities in the science data model (SDM) format were converted to measurement set (MS) format using the CASA tasks \texttt{importasdm}. The MS files were clipped for zeros using standard CASA flagging \texttt{flagdata}. We also applied \texttt{Hanning-Smoothing} and \texttt{tfcrop} to remove RFI from the data. We used the CASA calibration tables from the NRAO Archive for this observation and applied them to the raw data using the task \texttt{applycal}. The bright quasar 3C286 was used for flux and bandpass calibration. The phase calibrator J1911-2006 was observed for 90 seconds at regular intervals to calibrate complex gain fluctuations over time. The calibrated bursts were imaged using the CASA task \texttt{tclean} and the task \texttt{imfit} was used to fit an elliptical Gaussian to the burst in the image to get the centroid position, flux density, and the 1$\sigma$ image plane uncertainties.

To determine the accuracy of the astrometric reference frame in our VLA observation (and ultimately to report the burst position), we created a deep image using the whole VLA pointing on this field, and ran \textsc{PyBDSF}\footnote{https://github.com/lofar-astron/PyBDSF} to extract radio sources from it. This resulted in the detection of 99 radio sources. We then selected bright, compact radio sources using the following criteria: 1) The peak intensity per beam of the source (in Jy/beam) should be 0.7 times greater than the total integrated flux density of the source  (in Jy) in 1.5 GHz images, 2) the S/N of the source (ratio of peak intensity and the root-mean-square of the background) should be greater than 5. There were 23 radio sources after the cut-off was applied. The positions of the radio sources had an average image-plane statistical error as reported by \textsc{PyBDSF} of 0.06$''$ in RA and 0.11$''$ in DEC. 

We then cross-matched the radio point sources with the optical PAN-STARRS DR2 catalog, which is referenced to the GAIA2-based astrometric reference frame. We identified radio/optical associations searching for the nearest PAN-STARRS-DR2 source from each radio position, finding 23 optical counterparts in the PAN-STARRS-DR2 catalog with a maximum separation from the radio component of 2.72$''$. 

To determine whether there is a systematic offset between our radio imaging and the PAN-STARRS-DR2 source catalog, we subtracted the coordinates of the radio sources from the matched coordinates of the optical counterparts. We then averaged the offset values to determine a systematic relative offset of $\rm \Delta RA_{sys}=0.22''$ and $\rm \Delta DEC_{sys}=0.02''$. The standard deviation in $\rm \Delta RA_{sys}$ is 1.4$''$ and in $\rm \Delta DEC_{sys}$ is 1.18$''$. 

\begin{figure*}
    \centering
    \includegraphics[width=1\textwidth]{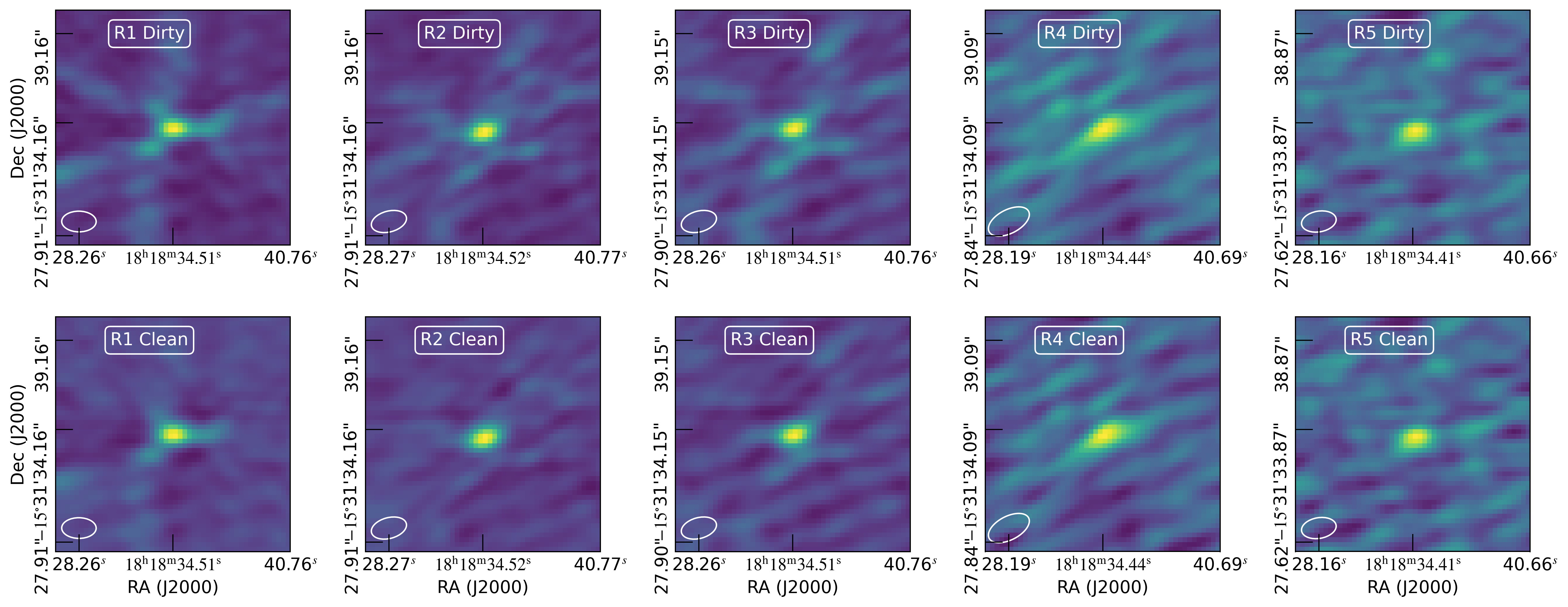}
    \caption{The dirty image (on the top panel) and clean image (on the bottom panel) of the \realfast\, bursts at 1.4 GHz. We have not used R4 and R5 for the positional analysis. The synthesized beam is shown in the lower left corner.}
    \label{fig:realfast}
\end{figure*}

The average burst position at L-Band is RA$=18:18:34.5206$ and DEC$=-15:31:34.1688$. The average image-plane statistical error on the burst positions is $\rm \Delta RA_{avg}=0.036''$ and $\rm \Delta DEC_{avg}=0.085''$ as reported by the CASA image-plane fitting function \textsc{imfit}. To represent the full positional error on the bursts, we add in quadrature sum these burst statistical errors with the PAN-STARRS offset, arriving at a final positional error of $\rm \Delta RA_{avg}=0.22''$ and $\rm \Delta DEC_{avg}=0.09  ''$.
The \realfast\ burst properties are reported in Table \ref{tab:realfastbursts} and the dirty and clean images of the visibility images of the bursts are given in Figure. \ref{fig:realfast}



\begin{deluxetable*}{cccccccc}
\label{tab:realfastbursts}




\tablecaption{\realfast\, detected bursts}
\setlength{\tabcolsep}{2pt}

\tablehead{\colhead{ID}&\colhead{RA} & \colhead{$\rm RA_{err}$} & \colhead{DEC}& \colhead{$\rm DEC_{err}$} & \colhead{DM} & \colhead{Flux} & \colhead{S/N} \\ 
 &\colhead{(deg)} & \colhead{($''$)} & \colhead{(deg)} & \colhead{($''$)} & \colhead{($\rm pc~cm^{-3}$)} &\colhead{(Jy)}& } 
\startdata
    R1& 274.64383 & 0.02& -15.52615 & 0.06 & 1021.2 & $0.221\pm0.003$ & 31.3\\ 
    R2& 274.64385 & 0.02 & -15.52616 & 0.04 & 1007.3 & $0.204\pm0.003$& 28.2\\ 
    R3& 274.64382 & 0.03 & -15.52615 & 0.05  & 1016.5 & $0.161\pm0.003$ & 23.1\\ 
    R4& 274.64351 & 0.04 & -15.52613 & 0.12 & 1030.3 & $0.050\pm0.001$ & 8.6\\ 
    R5& 274.64340 & 0.04 & -15.52607 & 0.10 & 1002.7 & $0.058\pm0.002$ & 8.4\\ 
\enddata

\tablecomments{DM is the detection DM reported during offline \realfast\ refinement process.\\
Flux as reported by CASA \texttt{imfit}.\\
\realfast\ searches an extremely large number of samples and has a standard Gaussian noise threshold of 7. The last two bursts in the table not used for positional analysis but are reported here because of their general proximity in sky position and DM.}



\end{deluxetable*}


\vspace{-20pt}
\subsection{Follow-up Observations}
\subsubsection{VLA/realfast}
To reduce the contributions from scattering and attempt to observe the intrinsic pulse structure, we carried out a higher-frequency observation, observing for three hours with the VLA observation at S-Band (2000-4000\, MHz frequency) in phased-array mode under program code 19B-313. We configured the observation to observe in 64 sub-bands each of 32\,MHz width, with 64 channels per sub-band, 100$\mu$s time resolution, recording dual polarization. The observation was during the VLA reconfiguration from A to D (A$\rightarrow$D)
\subsubsection{Green Bank Telescope Epoch 1: 1.5 GHz}
We carried out observations with the 100-m Robert C. Byrd Green Bank Telescope (GBT) in three epochs. The first epoch (AGBT 20A-420) was a 2\,h~46\,m observation on 20th March 2020 at 09:41:43 UTC. These observations used the L-Band receiver with a center frequency of 1.5~GHz and a bandwidth of 1500 MHz (4096 channels). We recorded data with the VEGAS pulsar backend in 8-bit format,  a sampling time of 87$\mu$s, and a channel frequency of 366 kHz. We opted to record only the total intensity data. 
\subsubsection{Green Bank Telescope Epoch 2L: 1.5 GHz}
We carried out a second set of GBT observations (AGBT 20B-407) on 6th August 2020 at 22:11:35.00 UTC for a total of 5 hours and 40 min on the source. This observation used the L-Band receiver with a center frequency of 1.5 GHz and a bandwidth of 800 MHz (4096 channels). We used the VEGAS pulsar backend in 8-bit format with a sampling time of 81$\mu$s, and a frequency resolution of 195 kHz. We recorded the Full Stokes data in the IQUV format.  A bright quasar, J1445+0958 was observed for flux calibration, and a 1-minute noise diode scan was done on-source for polarization calibration. Hereafter we refer to this as ``Epoch 2L.''
\subsubsection{Green Bank Telescope Epoch 2C: 6 GHz}
We did the third GBT observation (AGBT 20B-407) on 31st August 2020 at 22:25:21.00 UTC for a total of 5\,h\,30\,m on the source. We used the C-Band receiver with a center frequency of 6 GHz and a bandwidth of 4500 MHz (12288 channels). The data was recorded using the VEGAS pulsar backend in 8-bit format and had a sampling time of 87$\mu$s and a channel frequency resolution of 366 kHz. We recorded the Full Stokes data in the IQUV format.  A bright quasar J1445+0958 was observed in the ON and OFF positions for flux calibration, and a 1-minute noise diode scan was done on the source for polarization calibration. A test pulsar B1929+10 was observed for 5 min to verify the calibration. Hereafter we refer to this as ``Epoch 2C.''
\subsection{Single pulse searches}
\begin{figure*}
    \centering
    \includegraphics[width=1\textwidth]{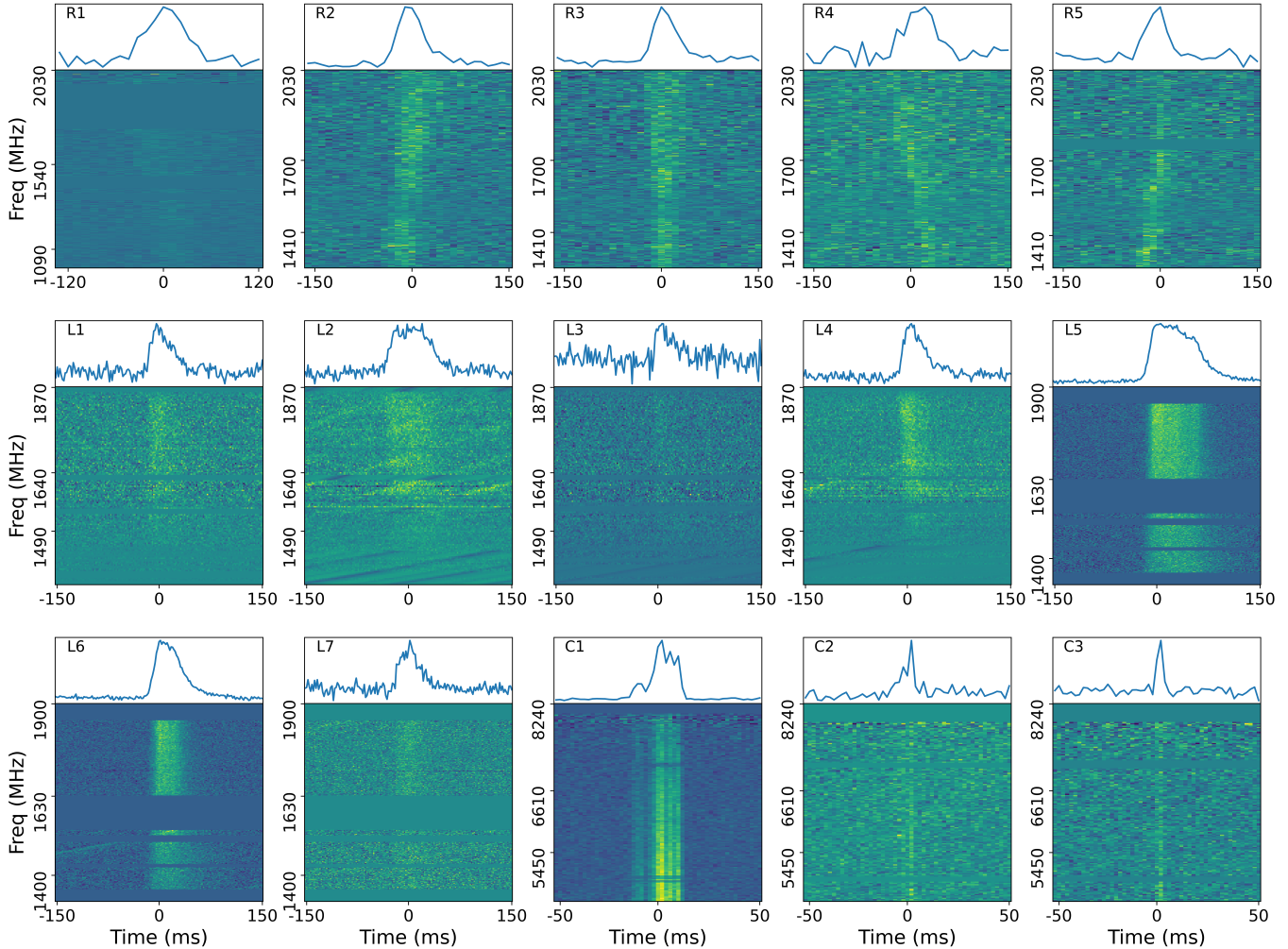}
    \caption{All the bursts detected by VLA/\textit{realfast} and GBT. The top panel shows the frequency-averaged, dedispersed time series of each bursts and the bottom panel shows the frequency-time spectrum of the dedispersed bursts. The \textit{realfast} bursts are dedispersed at the average DM = 1015.6 \dmunits. The \textit{realfast} data has a time resolution of 10 ms and is downsampled to a frequency resolution of 2 MHz. All GBT bursts are dedispersed at the average DM per epoch (ref. \ref{sec:fft}). The GBT burst data is downsampled to a time resolution of 1.95 ms and the frequency resolution of 1.95 MHz in L-Band and 23.42 MHz in C-Band. }
    \label{fig:spectrogram}
\end{figure*}

Here we describe the single pulse search done on the phased-array VLA S-band data and GBT data. The epoch 1 GBT data only had a usable bandwidth of 534 MHz, as the remaining parts of the band were either automatically filtered by the observing system or was manually flagged due to radio frequency interference (RFI). The GBT records data in \textsc{Psrfits} format, which we converted to \textsc{Filterbank} format using \texttt{your\_writer.py}\footnote{https://github.com/thepetabyteproject/your}. Custom RFI filters, with Savitzky–Golay (SG) and Spectral Kurtosis filter \citep{nita2010} with 4-$\sigma$ threshold and an SG filter window of 15 MHz were used during the writing process so that the converted \textsc{Filterbank} was RFI cleaned. The VLA data was already recorded in \textsc{Filterbank} format. Single pulse search was done using the python code \texttt{your\_heimdall.py}, which runs \textsc{Heimdall} \citep{barsdell2012heimdall} on the data, with a maximum boxcar width of 50 ms and a detection threshold of 6$\sigma$. The VLA data was dispersed at trial DMs between 990-1130\,\dmunits. The GBT epoch-1 data was searched in a wide range of DMs from 600-2500\,\dmunits. 
The Epoch 2L data was searched in a DM range of 600-2000 \dmunits, and the 6~GHz data was searched in a DM range of 800-1200 \dmunits. The Heimdall candidates were classified into real astrophysical signals both by the machine learning classifier \textsc{Fetch} \citep{Agarwal2020} and by visual inspection of the candidate plots.
A total of seven bursts were detected in the GBT data at 1.5~GHz, and three bursts were detected in the 6~GHz GBT data, above a signal-to-noise (S/N) ratio of 7. No bursts were detected in the phased VLA S-band data. Fig.~\ref{fig:spectrogram} shows the bursts detected during the GBT observations.



\begin{deluxetable*}{ccccccccc}
\label{tab:burst_properties2}



\tablecaption{\textbf{Properties of all the bursts detected by GBT}.\\
MJD is the arrival times referenced to infinite frequency and to barycentre.\\
S/N is the detection signal-to-noise reported by \textsc{Heimdall}.\\
DM is the signal-to-noise maximizing DM reported by \texttt{pdmp}.\\
$\tau$ is the scattering timescale.\\
$\chi^{2}_{red}$ is the reduced chi-squared value obtained for profile fitting.}


\tablehead{\colhead{ID} & \colhead{MJD} & \colhead{S/N} & \colhead{DM} & \colhead{Fluence} & \colhead{Width} & \colhead{$\tau$} & \colhead{$\chi^{2}_{red}$} \\ 
\colhead{} & \colhead{} & \colhead{} & \colhead{($\rm pc~cm^{3}$)} & \colhead{($\rm Jy\,ms$)} & \colhead{(ms)} & \colhead{(ms)} & \colhead{$\,$} } 

\startdata
L1 & 58930.418696580(8) & 7 & 1008.63$\pm$4.23 & 0.283$\pm$0.0528 & 16.3$\pm$1.96 & 25.4$\pm$2.17 & 1.05 \\ 
L2(a) & 58930.51873115(1) & 11 & 1021.46$\pm$3.87 & 0.701$\pm$0.139  & 16.4$\pm$2.87 & 26.8$\pm$2.87 & 1.35\\ 
L2(b) & 58930.51873150(2) &  &  &  & 24.2$\pm$6.54 &  \\
L3 & 58930.546648171(9) & 6 & 1002.29$\pm$3.62 & 0.172$\pm$0.0598 & 8.22$\pm$2.51 & 25.6$\pm$4.06 & 0.838\\ 
L4 & 58930.551574405(5) & 18 &  1011.86$\pm$2.71 & 0.662$\pm$0.0718  & 13.5$\pm$1.27 &   29.4$\pm$1.54&1.57 \\ 
L5(a) & 59087.980302006(6) &  &  & 2.30$\pm$0.327 & 8.66$\pm$1.91 &  \\
L5(b) & 59087.980301320(2) & 33 & 1013.75$\pm$0.37 & 25.7$\pm$0.879 & 52.8$\pm$1.76 & 18.7$\pm$0.722 & 1.30 \\
L5(c) & 59087.98030161(1) &  &  & 10.2$\pm$0.742 & 14.6$\pm$0.899 &  \\
L6(a) & 59088.014787158(5) &  &  & 1.65$\pm$0.726 & 3.26$\pm$1.81 &  \\
L6(b) & 59088.014787096(1) & 28 & 1011.39$\pm$0.37 & 12.5$\pm$2.28 & 18.1$\pm$1.79 & 18.5$\pm$0.826 & 1.08 \\
L6(c) &59088.01478733 (2) &  &  & 6.34$\pm$1.64 & 17.8$\pm$3.62 &  \\
L7 & 59088.052215423(1) & 7 & 1014.54$\pm$2.28 & 2.87$\pm$0.121 & 24.2$\pm$2.57 & 16.0$\pm$2.57 & 1.00 \\
C1(a) & 59093.035770803(1) &  &  & 0.0816$\pm$0.00377 & 4.76$\pm$0.305 &  \\
C1(b) & 59093.035770924(1) & 66 & 1003.68$\pm$ 4.23 & 0.334$\pm$0.00718 & 6.53$\pm$0.206 & 0.927$\pm$0.101 & 2.33 \\
C1(c) & 59093.0357709811(7) &  &  & 0.0562$\pm$0.00735 & 1.64$\pm$0.204 &  \\
C1(d) & 59093.0357710179(6) &  &  & 0.119$\pm$0.0058 & 2.51$\pm$0.266 &  \\
C2 &59093.120416971(2) & 9 & 1002.78$\pm$4.24  & 0.0296$\pm$0.00201 & 2.76$\pm$0.0958 &0.573$\pm$0.0105 & 0.947 \\
C3 & 59093.155093193(3) & 13 & 999.93$\pm$4.24 & 0.0273$\pm$0.00221 & 2.24$\pm$0.456 & 0.831$\pm$0.452 & 1.23 
\enddata




\end{deluxetable*}
\section{Periodicity Search} \label{sec:periodicity}
Given the millisecond-duration of the bursts, the simplest explanation of this source is that it is a neutron-star related phenomenon. Thus, we aim to determine whether there is any periodicity in the pulse arrival times. We carried out several periodicity search techniques.
\subsection{Fast-Fourier-transform (FFT) searches}\label{sec:fft}
We carried out FFT searches using the standard pulsar analysis package \textsc{PRESTO} \footnote{https://github.com/scottransom/presto}. As per standard \textsc{PRESTO} procedure, we mitigated RFI using the \texttt{rfifind} mask, and created dedispersed time series of the data using \texttt{prepdata} in the DM ranges 995-1025\,\dmunits. We ran FFT on the resulting time series to look for periodic signals using the code \texttt{realfft}, followed by \texttt{accelsearch}, which runs Fourier-domain acceleration searches and harmonic summing, with \textit{zmax}=200. We also ran the search without acceleration (\textit{zmax}=0). 
We then folded the time-series data at all candidate periods with a S/N $> 6$ using the function \texttt{prepfold}. 
There were multiple  S/N $> 6$ candidates in all searches. Later, the frequency-time \textsc{Filterbank} files at each epoch were dedispersed at their respective average single-pulse DM ($\rm DM_{avg}$: Epoch 1=1010~\dmunits, Epoch 2L =1012~\dmunits and Epoch 2C=1001~\dmunits), and folded at all the periods reported by the \texttt{accelsearch} above S/N$>6$. However, when the plots were visually inspected, it was apparent that all of these significant candidates were due to isolated instances of narrow-band RFI. The search was done in the barycenter frame of reference. 
\vspace{-3pt}

\subsection{Fast Folding Analysis}
We also carried out a fast-folding analysis (``FFA''), which allows superior sensitivity to long-period ($\gtrsim$1\,s) candidates than the FFT search. The minimum separation between two bursts (L3 and L4), 7.09 minutes, sets the upper limit to the possible period. Therefore we use 8 min (480 s) as the maximum period we search for in our trials. We ran the FFA software \textsc{riptide} \citep{Riptide} on the time series data produced from \textsc{presto} for all three epochs, dedispersed at the average DM at the respective epoch. We ran three separate searches on the time ranges from 0.01 - 1 s, 1 - 10 s and 10 - 480 s, analyzed the periodogram and folded the time series at the highest S/N period detection. The folded sub-integrations plot had no significant detections in all three epochs.

\subsection{Time-interval difference fitting}
This fitting analysis is commonly used to search for RRAT-like sources with a sparse number of detections. This technique seeks the largest integer division between the given time intervals. Below, we detail considerations taken when attempting such fits on this object and describe the outcome of the searches. 
Because some pulses had multiple peaks, and the epochs were widely spaced, here we identify distinct sets of pulses that we could search and how we time-tagged the bursts.

\subsubsection{Pulse selection for fitting}

\begin{figure*}
    \centering
    \includegraphics[width=1\textwidth]{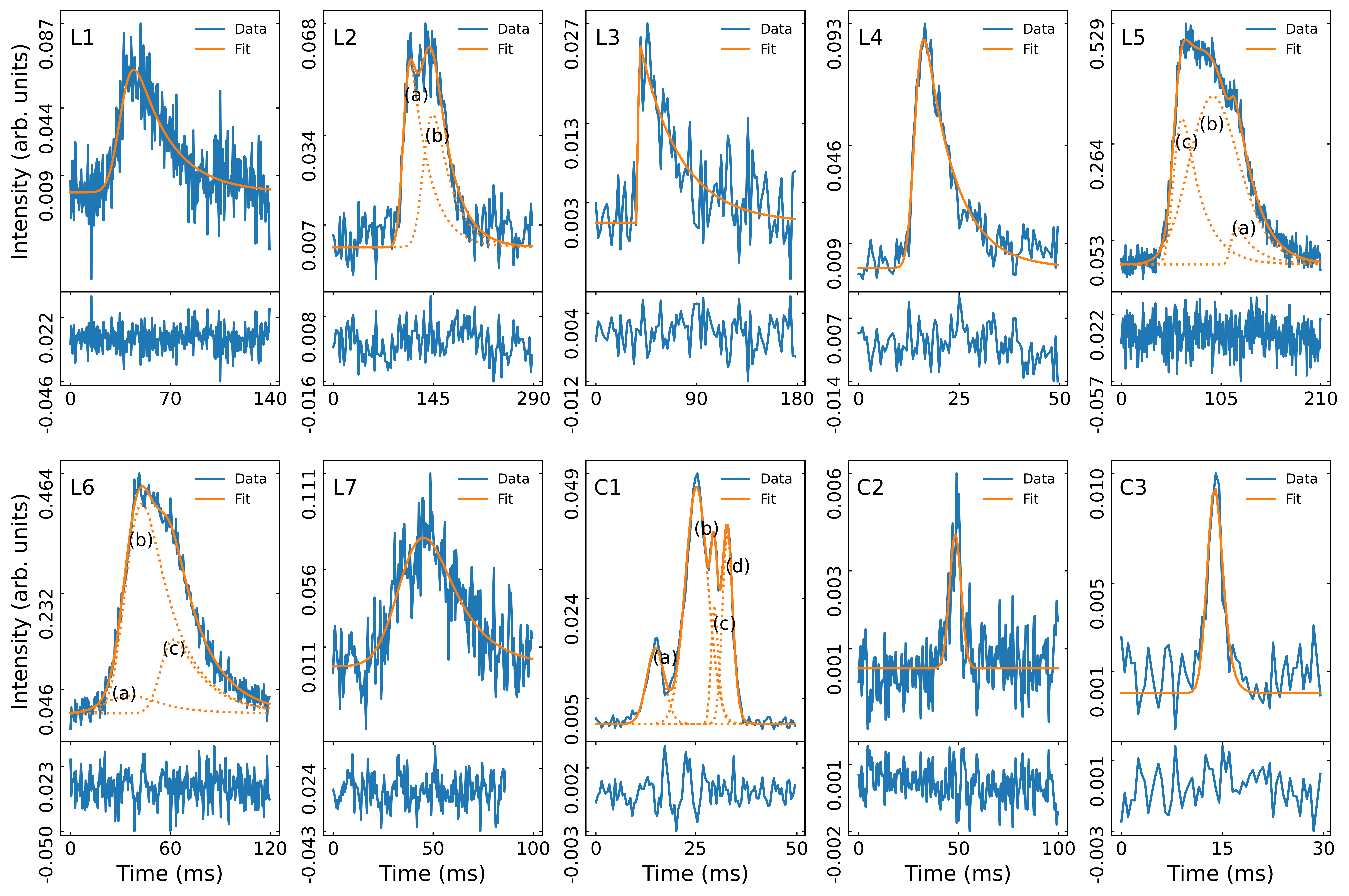}
    \caption{The profile of GBT bursts fitted by Gaussian convolved with exponential tail model. Some bursts are a sum of multiple Gaussian exponentials. The orange solid line shows the fit to the profile, and the dotted orange lines show the fitted components. The multiple components contribute to the sub-pulse sample. The fitting residuals are shown in the bottom panel.}
    \label{fig:fitting}
\end{figure*}

We detected 4, 3, and 3 distinct pulses at epochs 1, 2L, and 2C, respectively; we refer to these 10 bursts as the ``full-pulse sample''. We also detected what appear to be scattered sub-components of pulses. As these may or may not be due to separate rotations of a neutron star, we also carried out separate searches where we treated sub-pulses as distinct detections. This gives 5, 7, and 6 pulses in the epoch 1, 2L, and 2C data, respectively. This we will hereafter refer to as the ``sub-pulse sample''. These components or sub-pulses can be seen in Figure \ref{fig:fitting}.
\subsubsection{Pulse timestamp measurements}
Due to the complex burst morphologies, to fit properly for a periodicity the selection of how to quantify the burst arrival times is an influential component.

For the full-pulse sample, we used the average arrival time from the burst profile and used that as the input for the time-interval differencing. We calculated the MJDs for the full-pulse sample by averaging the sub-pulse MJDs for each bursts. 
For the sub-pulse sample, we use the mean of the fitted Gaussian for each sub-pulse as the time of arrival of the pulse. These MJDs are given in Table~\ref{tab:burst_properties2}. We ran the periodicity search codes on the following pulse subsets:
\begin{enumerate}
    \item All pulses (including all epochs and frequencies). We ran the search on both full-pulse and sub-pulse samples.
    \item Individual epochs treated separately. Only the sub-pulse sample was used in this trial.
    \item All L-Band MJDs (this avoids potential issues with burst morphology evolution with frequency; however, the period between epochs 1 and 2L may differ due to non-negligible $\dot P$ or errors in position). We only used the sub-pulse sample in this trial.
    \item All MJDs from epoch 2L and 2C to account for possible spin down. For a very large spin-down rate, $\rm \dot{P}=10^{-9}s s^{-1}$, the change in the period between the first and second epoch will be $\approx$13 ms. We only used the sub-pulse MJDs for this one.  
\end{enumerate}
We used the code \texttt{getper.py} \footnote{https://github.com/evanocathain/Useful\_RRAT\_stuff} to do this and searched in a range of periods from 0.01 - 480 s in different trials. The data was then folded at the first candidate period, which matched all the unique time differences from each set of trials. None of the trials yielded a detection. 


\section{Burst properties} \label{sec:properties}
We determined the fluence, width, arrival time MJDs, and scattering time of all GBT bursts using the software package \textsc{Burstfit} \citep{Aggarwal2021}, which uses \texttt{scipy.curve\_fit}. We averaged the data over the entire band and fitted the profile with a Gaussian convolved with an exponential scattering tail \citep{McKinnon2014} given by:

\begin{eqnarray}
\label{eq:1}
    P(t,\tau)=\frac{F}{2\tau}\exp\frac{\sigma^{2}}{2\tau^{2}}\times 
     \exp\left [-\frac{(t-\mu)}{\tau}\right]\\
     \times \left\{1+\rm erf\left[\frac{t-(\mu+\frac{\sigma^{2}}{\tau})}{\sigma\sqrt{2}}\right]\right\}       
\end{eqnarray}
where $F$, $\sigma$, $\mu$ is the area, standard deviation, and mean of the Gaussian pulse, and $\tau$ is the scattering tail. Some of the bursts were fitted with multiple components, if adding components resulted in a reduced chi-squared value closer to 1. The frequency channels that were zapped were removed before averaging. The MJDs reported are derived from the $\mu$ of the pulse and are referenced to infinite frequency and to the barycentre. We didn't have a flux calibrator for epoch 1 L-band bursts (L1--L4). Therefore, for that epoch only, the flux density is derived from the radiometer equation given by:
\begin{equation}
 S_{peak}=\frac{S/N~T_{sys}}{G\sqrt{n_{p}W\Delta f}}   
\end{equation}
where S/N is the signal-to-noise of the burst. $T_{sys}=20$K for GBT L-band receiver and $G$ is the gain which is equal to 2 K/Jy \citep{GBT}, $n_{p}$ is the number of polarizations which is equal to 2, $W$ is the width of the pulse and $\Delta f$ is the bandwidth of the burst which is equal to 500 MHz. The S/N is calculated using the equation:
\begin{equation}
    S/N= \frac{p_{max}-\Bar{p}}{\sigma_{p}}
\end{equation}
where $\sigma_{p}$ and $\Bar{p}$ is the off-pulse standard deviation and mean. 
This flux density was multiplied by the width of the burst (total width in case of L2) obtained from profile fitting to get the fluence.

We used the \textsc{psrchive} \citep{psrchive+hotan} package \texttt{pdmp} to optimize the DM that produced the highest S/N detection. The fluence is derived from $F$ in equation (\ref{eq:1}) for all the bursts in epoch 2, and the width is defined as the full-width-half-maximum of the Gaussian. The burst properties are given in Table \ref{tab:burst_properties2}. 
\subsection{Polarimetry}
\begin{figure*}
    \centering
    \includegraphics[scale=0.45]{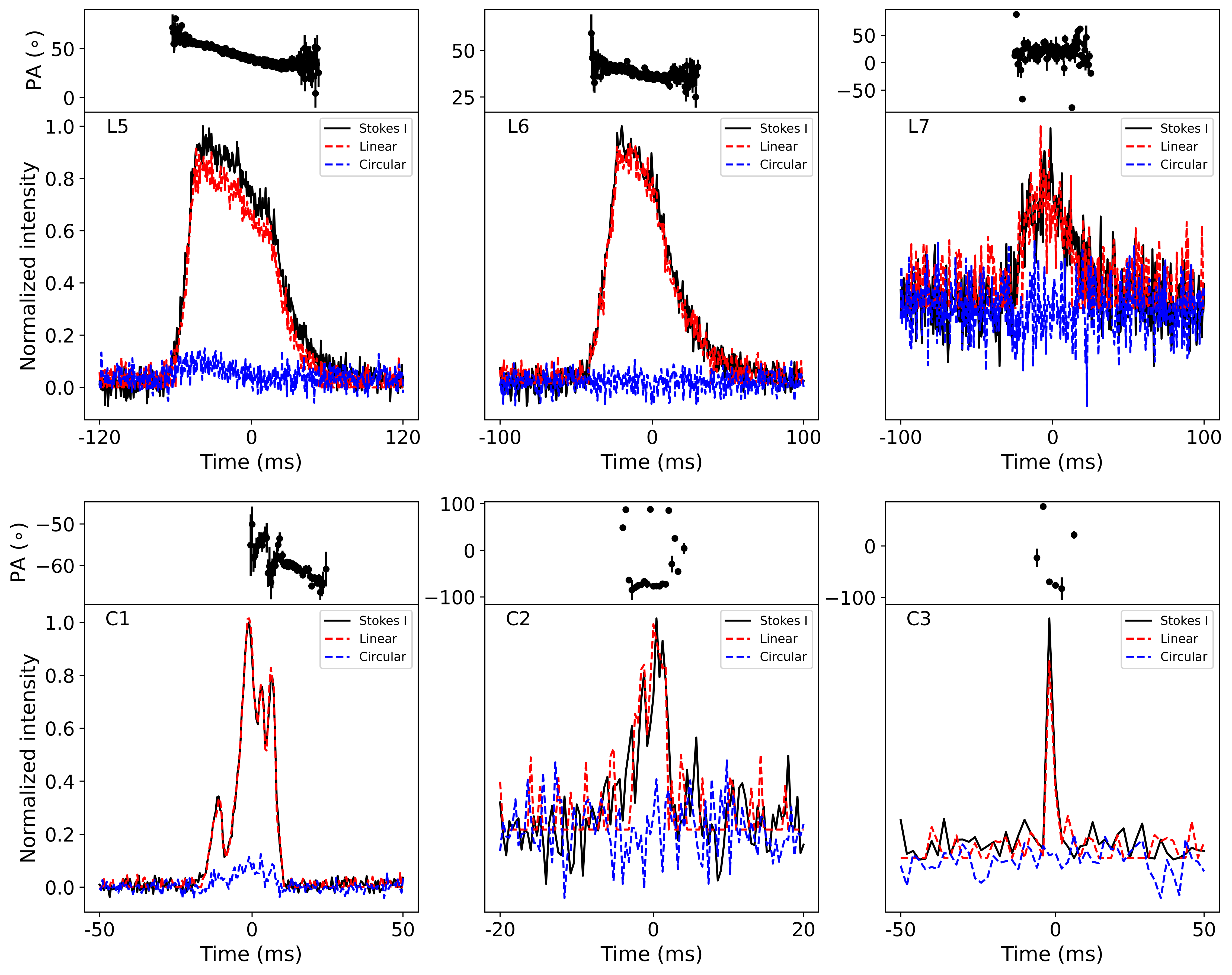}
    \caption{Normalized polarization profiles and PA for GBT bursts. The red and blue dashed lines indicate linear and circular polarization, respectively. The top panel shows PA curves fitted with 1$\sigma$ error bars. The burst names are also indicated.}
    \label{fig:polfigure}
\end{figure*}
Bursts in epoch 1 did not have a polarization calibrator available. For all epoch 2 bursts (L and C), we calibrated polarization using the standard package \texttt{pac}. We did 1-D RM Synthesis \citep{BrentjensBruyn2005,GHeald2009} on the Stokes data to search an RM range between $\pm10^{6}$ \rmunits and found an average RM of 816 \rmunits\ with a standard deviation of 9.82 \rmunits. We corrected the Faraday rotation by de-rotating the bursts at their respective RMs using the \textsc{psrchive} package \texttt{pam}. There were no significant peaks in the Faraday Dispersion Function of C3. Therefore, we de-rotated C3 at the mean RM of C1 and C2 i.e 828 \rmunits. The bursts were averaged in frequency, and the unbiased linear polarization was calculated using the following \citep{Everett2001}:
\begin{equation}
L_{\rm unbias} =
\left\{ 
  \begin{array}{ c l }
    \sigma_{I}\sqrt{\frac{L_{\rm meas}}{\sigma_{I}}-1} & \quad \textrm{if } \frac{L_{\rm meas}}{\sigma_{I}} \geq 1.57 \\
    0                 & \quad \textrm{otherwise}
  \end{array}
\right.
\end{equation}
where $\sigma_{I}$  is the off-pulse standard deviation in Stokes I and $L_{\rm meas} = \sqrt{Q^{2}+U^{2}}$. The linear polarization fraction was calculated using:
\begin{equation}
    \frac{L}{I} = \frac{\int L_{unbias}~dt}{\int I~dt} \
\end{equation}
\begin{equation}
    \sigma_{L/I} = \frac{L}{I}~\sqrt{\left( \frac{\sigma_{L_{unbias}}}{\int L_{unbias}~dt}\right)^{2}+\left(\frac{\sigma_{I}}{\int I~dt}\right)^{2}}
\end{equation}
The circular polarization fraction (V/I) was also calculated similarly and the absolute value of V/I is reported.
The polarization angle (PA) and its error was calculated using:
\begin{equation}
    \phi_{0}=\frac{1}{2} tan^{-1}\frac{U}{Q} 
\end{equation}
\begin{equation}
    \sigma_{\phi}=28.65^\circ\frac{\sigma_{I}}{L_{unbias}}
\end{equation}
All the bursts are $>$ 85\% linearly polarized, and up to 12\% circularly polarized. The PA shows a gradual change in some bursts, and some bursts show nearly flat polarization angles throughout the duration. In pulsar studies, flat PA arises because of these reasons: 1) Scattering smearing \citep{Li2003}; 
2) Our line-of-sight grazing the edge of the emission cone; or 3) Nearly aligned magnetic and spin axis \citep{Hurley-Walker2022}. 
The polarization properties of the bursts are given in Table. \ref{tab:poltable} and the polarization profiles are illustrated in Figure. \ref{fig:polfigure}.




\begin{deluxetable}{cccc}




\tablecaption{Polarization properties of the bursts}

\label{tab:poltable}
\tablehead{\colhead{ID} & \colhead{RM} & \colhead{L/I} & \colhead{$\rm \mid V/I\mid$} \\ 
\colhead{} & \colhead{($\rm rad~m^{2}$)} & \colhead{} & \colhead{} } 

\startdata
L5 & 807.70$\pm$3.17 & 0.85$\pm$0.000006 & 0.12$\pm$0.00006 \\
L6 & 808.92$\pm$0.74 & 0.93$\pm$0.0001 & 0.02$\pm$0.0001 \\
L7 & 811.61$\pm$1.71 & 0.98$\pm$0.0007 & 0.07$\pm$0.0008 \\
C1 & 833.24$\pm$11.2 & 1.00$\pm$0.0001 & 0.09$\pm$0.0001 \\
C2 & 823.16$\pm$26.1 & 1.00$\pm$0.004 & 0.005$\pm$0.004 \\
C3 &   -             & 0.82$\pm$0.04  & 0.12$\pm$0.04 \\
\enddata




\end{deluxetable}

\vspace{-2pt}
 \subsection{Spectral index and Energetics}\label{sec:spix}
\begin{figure}
    \centering
    \includegraphics[width=0.5\textwidth]{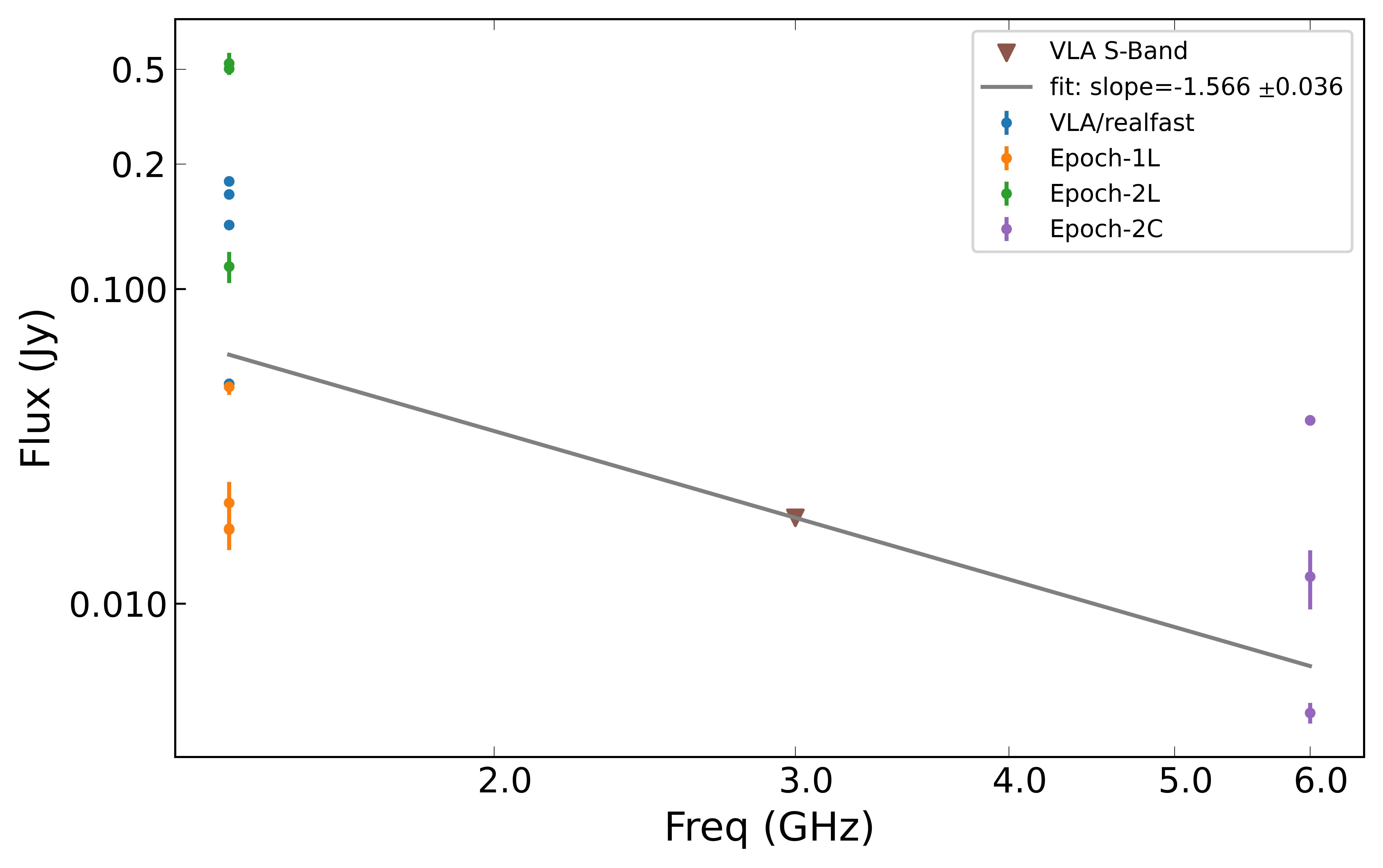}
        \caption{The burst fluxes as a function of frequency. The dots shows the flux in Jy. The error bar measurement is described in  \S\ref{sec:spix}. The grey line shows the best-fit slope to all data points as described in \S\ref{sec:spix}}. 
    \label{fig:freqspectrum}
\end{figure}

 The detection of bursts at different frequencies ranging from 1.1--1.9~MHz and 4--8~GHz allows us to calculate the spectral index for the source. Since the burst detections are non contemporaneous and has a large gap between them, we have to make the assumption that the source is stable during this timescale and does not show any intrinsic flux variability at a given frequency.
 All the 15 bursts from the VLA/\realfast and GBT Epoch 1 and Epoch 2L and 2C are plotted in Figure. \ref{fig:freqspectrum}. The 8$\sigma$ value of VLA S-band is also included for completeness. The flux of each burst in epoch 2L and epoch 2C, was calculated by dividing the total fluence of each burst by the total width. The error on the flux is calculated by propagating the errors on fluence and width. The flux of Epoch-1 L-band bursts are calculated as described in \S\ref{sec:properties} and the error bar is calculated from the radiometer noise. Assuming a power law variation of flux with frequency, $S\propto f^{\alpha}$, we fitted a curve with two free parameters given by:
 \begin{equation}
     S = S_{0}f^{\alpha}
 \end{equation}
 where $\rm S_{0}$ is the y-axis intercept and $\alpha$ is the power-law index, using \texttt{scipy.optimize.curvefit}. The error bars on the fluxes were accounted for while fitting the curve. Here, the slope of the curve in the log scale is equivalent to the spectral index $\alpha$ of the source. The source has a spectral index of --1.5$\pm$0.036. 
 
\begin{figure*}
     \includegraphics[width=1\textwidth]{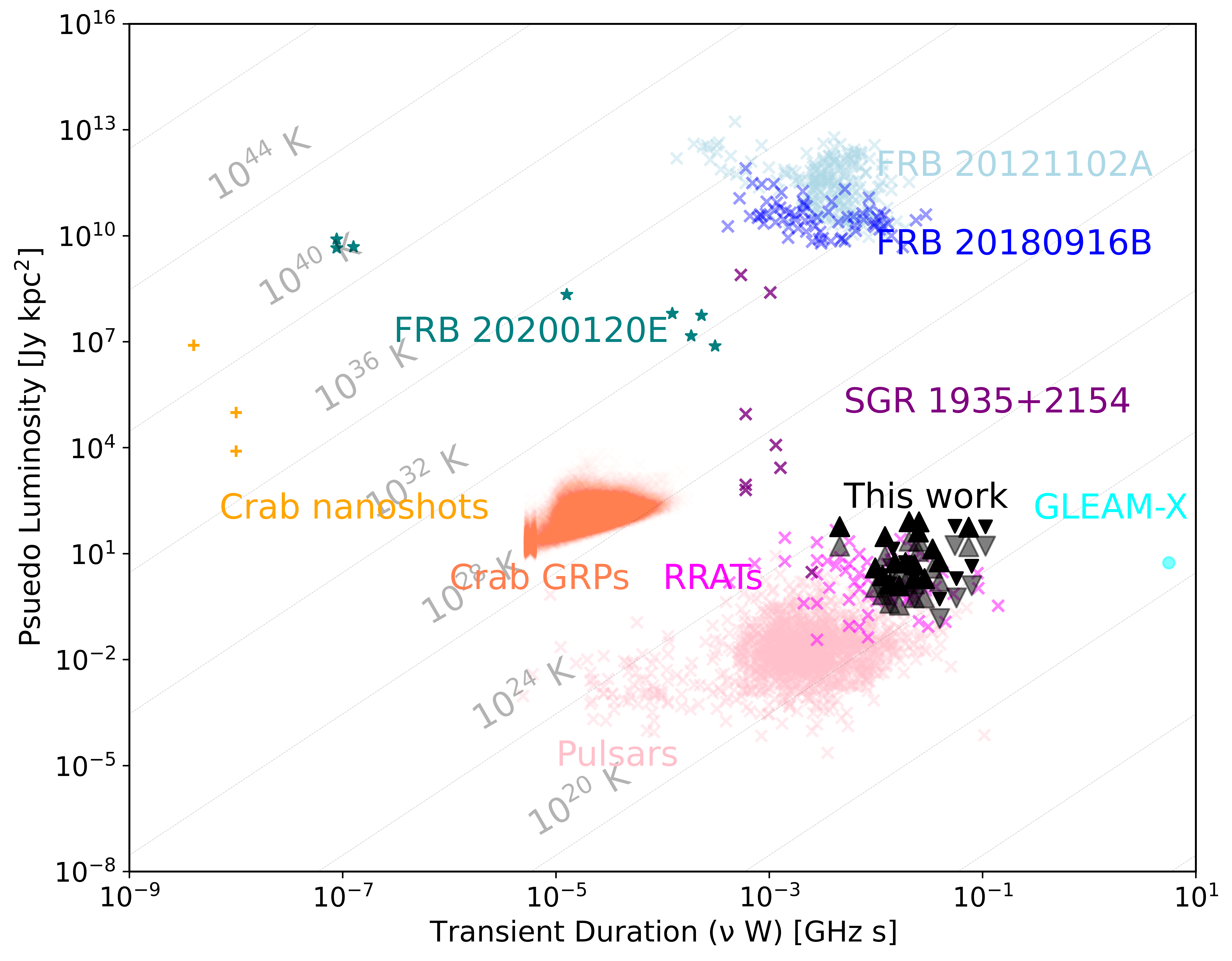}
     \caption{Transient phase space. The black triangles shows bursts from J1818--1531. The downward triangles represent the full pulse sample, and the upward triangle represents the sub-pulse sample. The darker and lighter shades of the triangles represent the luminosities using NE2001 and YMW16, respectively. This figure is adapted from \citep{Nimmo2022}}
     \label{fig:tps}
 \end{figure*}

 Using the DM-distance, we also calculated the isotropic luminosity of the source at 1.4~GHz given by $L=4\pi d^{2}S$. 
The maximum isotropic luminosity of all the bursts in the full-pulse sample is $\rm 7.4\times10^{22} ~erg~s^{-1}Hz^{-1}$ ($\rm 2.09\times10^{22} ~erg~s^{-1}Hz^{-1}$) using d=10.9 kpc (5.8 kpc), from burst L6. We also calculated the pseudo luminosity given by $L_{\nu}=S_{\nu}d^{2}$ for the full pulse and sub-pulse sample is plotted in Fig. \ref{fig:tps} with other fast radio transient sources. We can calculate the brightness temperature of the source using
 \begin{equation}
     T_{B} = \frac{S_{peak}}{2\pi k_{B}}\left (\frac{f\Delta t}{d}\right)^{-2}
 \end{equation}
The brightness temperature distribution for all the bursts components has a mean value of $2.23(0.644)\times10^{23}$ and a 1$\sigma$ value of $6.77 (1.95)\times10^{23}$ for NE2001(YMW16) distances. Note that we removed burst L2 from this calculation to avoid any inconsistencies.

\subsection{Scattering}\label{sec:scattering}
\begin{figure}
    \centering
    \includegraphics[width=0.5\textwidth]{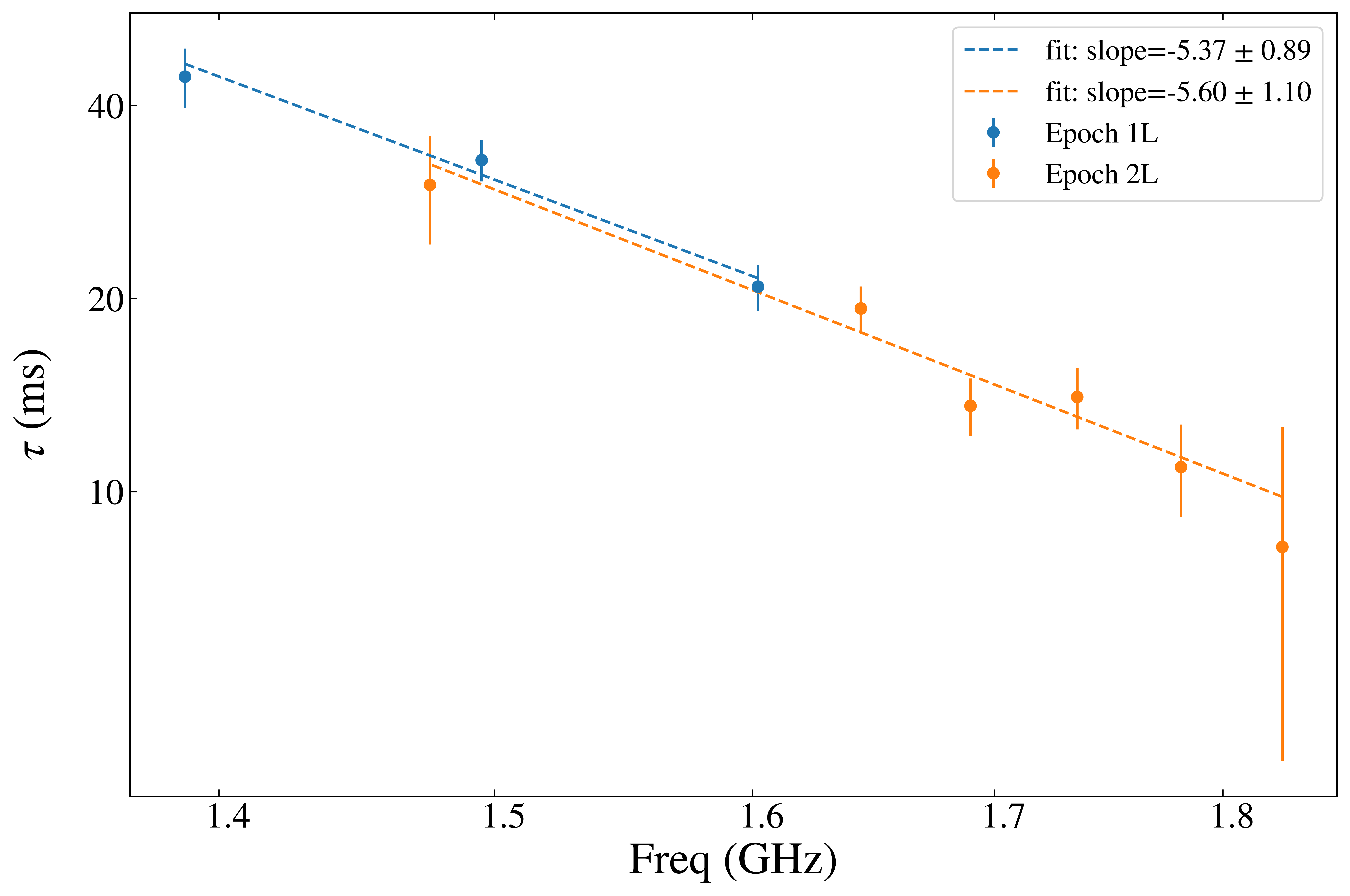}
    \caption{The frequency dependence of scattering timescales for bursts in epoch 1 and 2 L-Band. Burst L4 (in blue) from epoch 1 and burst L6 ( in orange) from epoch 2 is plotted and their best fitting curve is marked by dashed lines of respective colors. See \S\ref{sec:scattering} for more details about the plot.}
    \label{fig:tau}
\end{figure}

 The scattering timescale was measured by assuming that the burst profile was a Gaussian convolved with a one-sided exponential tail with a $1/e$ delay given by $\tau$. 
The bursts in our sample show scattering at both L and C bands, as reported in Table~\ref{tab:burst_properties2}.  There is a noticeable difference in the scattering timescale between the two epochs of 1.4 GHz observations with a mean value of 26.56$\pm$1.40 ms in epoch 1 and 17.63$\pm$0.69 ms in epoch 2. This implies a variable scattering screen between the source and the observer. The scattering timescale reported for C-Band bursts could be mischaracterization due to complex burst morphology. To test the credibility of the scattering timescale variation between epochs, we plotted the scattering timescale vs the frequency for the bursts L4 from Epoch 1 and burst L6 from Epoch 2. The bursts were selected based on the criteria that they have enough S/N to be split into different bands and they have good model fit. The sub-bands with poor fit were removed. Three sub-bands of L4 and 6 sub-bands of L6 are used. The data was fit with $\tau \propto f^{\alpha}$. The best fit curve for L4  has a $\alpha=-5.3\pm0.89$ and L6 has a $\alpha=-5.6\pm1.1$ (Fig. \ref{fig:tau}). For Kolmogorov spectrum, $\tau \propto f^{-4.4}$. The index of scattering timescales in both epochs are consistent within error bars with that of a Kolmogorov medium and therefore suggests that the epoch-to-epoch variability seen in L-band bursts are indeed real.
\section{Discussion} \label{sec:discussion}
First, we present a summary of basic observations. The location of the source towards the Galactic plane explains the high DM, and the source is decidedly of Galactic origin. The smallest burst duration of 1.6 ms suggests that the object may have a radius of a few hundred kilometers. The high brightness temperature of $\sim 10^{23}$K suggests a coherent emission mechanism. 
The high fraction of linear polarization implies ordered magnetic fields.
The bursts show stochastic variations in DM around a mean of 1009 \dmunits and a standard deviation of 6 \dmunits.
The scattering timescale decreased by about $\approx$ 40\% between the first and second epoch L-Band observation, suggesting a dynamic scattering screen. This variation appears to be uncorrelated with DM variations. The relatively low luminosity and, accordingly, low brightness temperatures argue away from a Galactic-FRB scenario for the pulses for now, even though more energetic bursts similar to the ones seen from SGR 1935+2154 cannot be excluded to occur in the future.

The generic properties as described above point decidedly towards a neutron-star-related phenomenon. The lack of periodicity detection in these pulses does not rule out that scenario, as some periodicities can be difficult to identify in sparse pulse samples (e.g. mode-changing RRATs, RRATs with a wide duty cycle, scattered MSPs, or magnetars with a $\sim$100\% duty cycle; e.g.~\citealt{bsb10,Levin2012,sun+21}). While this does present challenges in the characterization of this source, below we compare the properties of the bursts with known neutron-star populations to identify any common behavior.

\begin{figure*}
    \centering
    \includegraphics[width=0.7\textwidth]{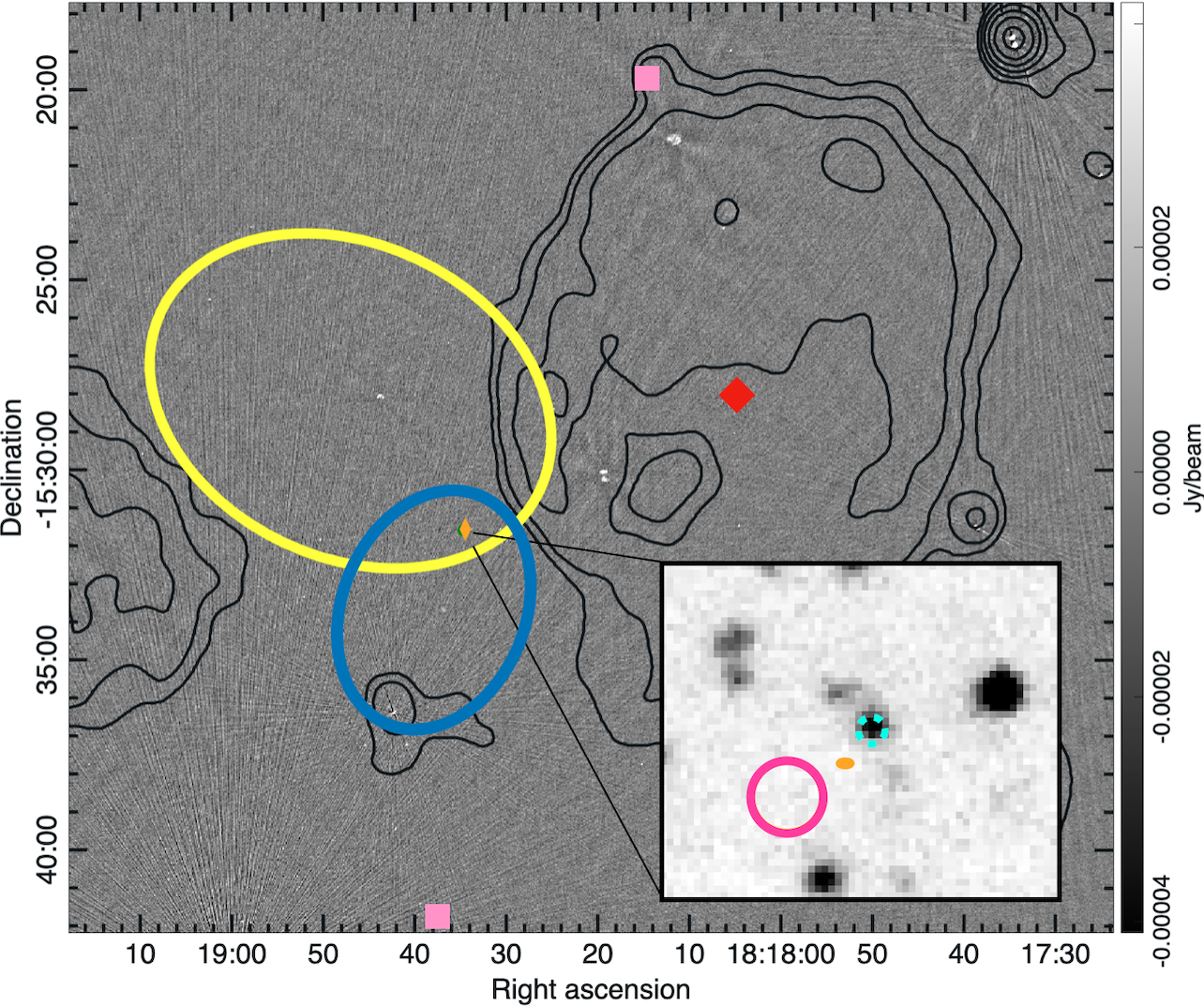}
\vspace{-5mm}
    \caption{Here we display the relative layout of field sources and error regions. The main image shows the deep VLA observation in greyscale. The large yellow and small dark blue ellipses represent the 3FGL and 4FGL \textit{Fermi} detections, respectively. The orange marker represents the \realfast\ localization. The light pink markers in the northern and southern edge of the image are unrelated pulsars. The black contours show 330\,MHz emission in the region from \citet{brogan+06} at levels at $9\,{\rm mJy}\times 2^n$, where $n=(0, 1, 2, ...)$. The red diamond represents the HESS source, presumed PWN, associated with the supernova shown.
    The inset image shows the average \realfast\ RA/Dec errors represented as an ellipse (orange) on the greyscale PAN-STARRS i-band image, with the dotted blue circle representing the location and a nominal 0.5$''$ RA/Dec error radius of PSO~J274.6436-15.5258 (see Sec.~\ref{sec:field}). 
    The magenta ellipse to the southeast represents the position and error radius of the X-ray source identified by XMM-Newton.}
    \label{fig:fieldimage}
\end{figure*}

\subsection{Sources in the field and chance coincidence.}\label{sec:field}
%
%
We examined other observations of the field to determine whether any objects might be co-located or associated with the bursts at the \realfast\ position. 
Figure \ref{fig:fieldimage} summarizes the multi-frequency emission within approximately 0.25$^\circ$ of the bursts.
Ultimately, we come to the conclusion that while there is a star near the positional localization error region, there are no confidently associated stars, pulsars, or supernova remnants with the burst source. The \textit{Fermi} source, however, does appear to be confidently associated with the bursts.
\subsubsection{Co-located stars and pulsars}\label{sec:coloc}
There are four known pulsars within 0.25 degrees, as listed in the ATNF Pulsar Database \citep{psrcat}, however these appear to be unrelated given their much different positions and DMs ($<850$\,\dmunits). The nearest of these is still approximately 10 arcminutes away from the \realfast\ position, thus, none are spatially coincident.

A search of SIMBAD \citep{SIMBAD} and the NASA Extragalactic Database returned no objects within three times the RMS \realfast\ position error ($\Delta$RA and $\Delta$Dec  of 0.22$''$, $0.09''$); the nearest cataloged object was what appears to be an unrelated 2MASS source 5$''$ to the south-east. To check for any faint uncataloged stars co-located with the bursts, we searched the PAN-STARRS DR2 catalog within a 2$''$ radius. This returned one object ID within this region. The object, PSO~J274.6436-15.5258 (object ID 89362746437589401) has a reported DR2 stacked-image position (J2000 ${\rm RA}=274.64358\pm0.00100^\circ$, ${\rm Dec}=-15.52578\pm0.00100^\circ$), and is the bright compact source 1.58$''$ north-west from the \realfast\ position shown on the inset of Fig.~\ref{fig:fieldimage}. This source is consistent with the \realfast\ position within the error region quoted on the Pan-STARRS source. The source is identified by the GAIA catalogue as being of a Star object type with a 99.95\% probability. This is consistent with the PAN-STARRS measurements using the classification scheme set by \citep{tachibana-panstarrs-ident}, where we can obtain a crude object type identification (star vs. galaxy) by comparing the difference between the reported i-band PSFMag and KronMag values (computed as iMeanPSFMag--iMeanKronMag) in the PAN-STARRS catalog. The iMeanPSFMag value of 20.283 and difference value of --0.07 imply that this object is more likely to be a star than a galaxy. Using the iKronMag, zKronMag and the yKronMag values, we computed the i--z difference and the z--y difference. The differences were 0.91 and 0.47, respectively, which makes it consistent with the star having the spectral classification M5V and mass 0.16$\rm M_{\odot}$ \citep{Pecaut2013}. Given the proximity of this field to the Galactic center, it is expected that some spatially coincident emissions may arise by chance. Given the lack of spatial co-location of the burst source and this star, it is unlikely that these two objects are of a common origin.

We also queried NASA's The High Energy Astrophysics Science Archive Research Center (HEASARC) within a radius of 2$''$ to search for any high energy counterpart. The query returned an XMM Newton Source at J2000 RA and Dec 18:18:34.639, --15:31:35.1 with an error radius of 1.316$''$. It has a combined EPIC band 8 flux of $\rm (1.97\pm0.72)\times10^{-14} ergs~cm^{-2}~s^{-1}$. For a distance of 10.9 kpc, this would become an X-ray luminosity of $\rm 2.22\times10^{31}~ergs~s^{-1}$. The position of this X-ray source is not consistent with the \realfast\ position within the formal error, but is separated at a distance within 2$\sigma$ of the positional error.


%
%

\subsubsection{Supernova Remnant and Pulsar Wind Nebula}
%

Given the suspected association of our burst source and the \textit{Fermi} emission with a neutron star, we searched the broader field for evidence of nearby supernovae. To the west of the burst position and \textit{Fermi} regions is a composite supernova remnant, ${\rm G}15.4+0.1$, which contains a compact X- and gamma-ray PWN near its center (HESS J1818-154, e.g. \citealt{hess}). No pulsar has yet been identified as coincident with the PWN.

It does not appear that our burst source is a pulsar associated with this PWN or supernova event, based simply on its large offset; if it is associated with the PWN, the pulsar (or \textit{Fermi} emission) should be directly coincident with that emission. Neither the \textit{Fermi} or \realfast\ error regions are coincident with the PWN emission. The \realfast\ position sits at a distance of 6.1$'$ from the central position of the PWN. At the distance of 9.3\,kpc indicated by \citet{su}, and their modelled supernova age of 11\,kyr, the pulsar would have had to travel at a velocity of $>$1400\,km\,s$^{-1}$ to reach its current position (this minimum assumes the trajectory is perfectly orthogonal to our line-of-sight). This is much larger than typical pulsar birth velocities in the Milky Way \citep{lbh+97}. While there is some precedent for such a configuration \citep{schinzel+19}, the lack of morphological evidence of such a high velocity (trail or bow shock) in the VLA and archival data, in addition to the central PWN, argue against that scenario.

We also note that in the 4FGL catalog, 4FGL J1818.6-1533 appears to be listed as associated with SNR G015.4+00.1, likely due to its relative proximity. However, based on the positioning in Fig.~\ref{fig:fieldimage}, it appears the \textit{Fermi} source is more likely not associated with the remnant.


\subsubsection{Statistical association of the \textit{Fermi} error regions}

Here we present some general arguments to better understand whether the burst source is in fact associated with the \textit{Fermi} source. 

First, it is clear that the \textit{Fermi} 3FGL position is slightly different from that of 4FGL. Both 95\% \textit{Fermi} error regions are consistent with the \realfast\ bursts. The 4FGL position has a much smaller error region, and we characterize a conservative likelihood of chance coincidence for this occurring by answering the question, ``given a random \realfast\ position, what's the chance it would have landed within a random 4FGL error region of this size?'' We do this by estimating the density of \textit{Fermi} sources (number of sources per square degree) near the Galactic center region, then multiplying this by the angular coverage of the 4FGL region in Figure~\ref{fig:fieldimage}.
In the 4FGL catalog, we find 559 sources within the region covering $\pm20^\circ$ in Galactic longitude and $\pm10^\circ$ in Galactic latitude around the Galactic center. This tells us the source density per sky area is around 0.7 sources per square degree. Multiplying that by the area of the 4FGL ellipse (minor axis 144.36$''$, major axis 194.76$''$), we find a probability of 0.0048 that the \realfast\ position would overlap the 4FGL error region by chance. This number would be less if we considered only unidentified \textit{Fermi} sources, and took into account the additional confidence afforded by the 3FGL error region; thus, we consider the co-location, and likely association of the \textit{Fermi} source with this burst emitter, to be relatively secure.

If the source were associated with a $\gamma$-ray pulsar then following equation 24 of the \citet{Smith2023}, $L_{\gamma}=4\pi d^{2}f_{\Omega}G_{100}$, using the DM distance of 10.9 kpc and a phase-averaged integral energy flux in the 0.1 to 300 GeV energy band $G_{100}$ = $5.19\cdot10^{-11}\,$erg\,cm$^{-2}$\,s$^{-1}$, which yields a $\gamma$-ray luminosity of $7.25\cdot10^{35}$\,erg\,s$^{-1}$ with the beaming fraction, $f_\Omega = 1$. If the nature of the detected pulses were from a pulsar-like object, this would put it into the category of young pulsars.

\subsection{Likely sources of burst emission}\label{sec:hypotheses}
Assuming J1818--1531 is associated with the \textit{Fermi} emission, and taking into consideration the duration and coherent emission, the source could belong to one of the following:
\subsubsection{Magnetars}
Magnetars are a class of rotating neutron stars with very high magnetic fields, $\rm B_{surf} > 10^{14} G$ and belong to the younger class of neutron stars. They can exhibit a rotation period of 1-10 s and very high spin down up to $\rm 10^{-9}\,ss^{-1}$. They typically show time-variable flux densities, high linear polarization, profile changes, and flat spectral index \citep{Levin2012}. Some magnetars are known to show an inverted spectrum \citep{Camillo2007a,Levin2012}. 
Some magnetars also show circular polarization \citep{Levin2012,Kramer2007}. These are also known $\gamma$-ray and X-ray emitters in keV and MeV ranges. Their decaying magnetic field powers their emission, and they are capable of emitting radio bursts of energies similar to extragalactic FRBs, up to $10^{35}~ergs~s^{-1}$\citep{Bochanek2020,CHIME1935}. 

The high polarization fraction, variable fluxes, and energies of J1818--1531 are consistent with a magnetar origin. Magnetars are also known to have a large duty cycle up to 50\% \citep{Levin2012}, which can lead to a non-detection of a period. The gradual change or flatness of PA is also commonly seen in magnetars \citep{Camillo2007}. 

Magnetars have not been detected in GeV energies, except for one extragalactic magnetar giant flare detected by \textit{Fermi}-LAT \citep{Ajello2021}. However, that flare was transient (fading on a timescale of less than 300\,seconds). The high energy magnetar thermal emission typically arises in the X-ray band and is modelled by a blackbody in the soft X-ray band and a power law component in the hard X-rays. \cite{Takata2013} predicts non-thermal $\gamma$-ray luminosity from magnetars of the order $\rm \sim 10^{35}~ergs~s^{-1}$ in the GeV energies and suggests that the GeV photons can only escape the magnetar surface if the corresponding X-ray luminosity is $\rm <~10^{35}ergs~s^{-1}$. If the XMM Newton source is associated with J1818--1531, then the X-ray and $\gamma$-ray energies are consistent with this model and this will be the first detection of GeV energies from a Galactic magnetar. However, the offset between the X-ray and the radio sources currently disfavors this association.

\subsubsection{Millisecond Pulsar}\label{sec:msp}
Millisecond pulsars are recycled pulsars that, as the name suggests, have a period of a few milliseconds. They have larger characteristic age and, therefore, much weaker surface magnetic field strength. Many \textit{Fermi} $\gamma$-ray associated sources have been identified to be MSPs \citep{Ajello2021}. 
The bursts reported in our work could be bright single pulses from an MSP, while the low luminosity pulses could be suppressed due to scattering smearing. MSP single pulses are also highly variable, which is consistent with our observations. The high linear polarization fraction is also consistent with single pulses from some MSPs \citep{De2016}. Fractional circular polarization is also seen in MSP single pulses. However, the energies of J1818--1531 are approximately three order of magnitude greater than observed from MSP single pulses \citep{Palliyaguru2023}.  

The seemingly distinct pulse subcomponents in C-band, which appear to be wide intrinsically rather than due to scattering (as argued by their symmetry), seem to argue against millisecond-duration periods less than about 6.5\,ms, which is the longest sub-pulse duration we quantified (pulse C1b). This and the significant variability and $\gamma$-ray luminosities present some arguments against an MSP as the burst emitter.

\subsubsection{Young Pulsars}
Another class of $\gamma$-ray emitting neutron stars is young pulsars with periods between from fifty to a few hundred milliseconds. These can be prominent giant-pulse emitters. Giant pulses are very narrow pulses with widths ranging from nanoseconds to microseconds. Some studies have shown that young pulsars are more than 70\% linearly polarized. Circular polarization has also been detected in single pulses of young pulsars \citep{Kramer2002}. No obvious orthogonal jumps were reported in most and most of them also show flat PA swings \citep{Johnston2006}. Young pulsars are capable of emitting very bright single pulses too \citep{Large1968}. The polarization fraction, non-detection of PA jumps, brightness of single pulses, and $\gamma$-ray luminosities are consistent with the source being a young pulsar.

\subsubsection{Rotating Radio Transients}
Rotating Radio Transients (RRATs) are sparsely emitting, characterized by their detection more prominently in single-pulse rather than Fourier-based searches \citep{Mclaughlin2006,Burke-Spolaor2013}. They have spin periods ranging 45ms -- 7.7s and surface magnetic fields in the range $1.67\times10^{11}$ --$4.96\times10^{13}$ G \citep{Abhishek2022}. One RRAT has been previously detected in X-rays \citep{Rea2009}, but none has been associated with $\gamma$-rays yet. RRATs show diverse polarization properties, with some single pulses having 100\% linear polarization fraction and a significant amount of circular polarization \citep{Hsu2023}. Some RRATs also show discontinuous PA jumps as expected from orthogonal polarization modes \citep{Caleb2019}, which we do not see in our source. However, in Fig. \ref{fig:tps} the bursts from J1818--1531 occupy the same spaces as RRATs and if this is indeed an RRAT, it would be the first one to be discovered to have a $\gamma-$ray counterpart.

The absence of discontinuous PA jumps in the radio burst profile and the expected lack of $\gamma$-ray emission from an old pulsar makes it unlikely for J1818--1531 to be an RRAT.

\subsubsection{Binary star system}
While directly associating with the star described in  \S\ref{sec:field} proved unlikely, it is worth considering the implications if the source were in a binary. If there is somehow (e.g. due to unrecognized localization error issues) and association between the {\it realfast} bursts and the star discussed in \S\ref{sec:coloc}, the M-dwarf nature of the star will make the system a low-mass binary. $\gamma$-ray emission from low mass binaries suggests that the neutron star is a recycled MSP. But we disfavor an MSP as the source because of the reasons outlined in \S\ref{sec:msp}. However, GeV emissions have been detected from high-mass binaries like PSR B1259--63 and PSR J2032+4127. The GeV emission is suggested to arise from a combination of Bremmsstrahlung and inverse Compton emission from unshocked and weakly shocked electrons of the pulsar wind \citep{Chang2021}. The isotropic $\gamma$-ray luminosity during the periastron of PSR B1259--63 is consistent with the observed \textit{Fermi} luminosity of J1818--1531 \citep{Abdo2011}. A hidden high-mass companion star would be required to explain the GeV emission of J1818--1531. Eclipses by the companion star or absorption of the radio emission from the pulsar by the outflows from the star \citep{Chen2021} could also help explain the intermittency of pulses seen from this object, and a line-of-sight through a variable plasma medium (as during orbit) could provide the DM and scattering variations we observed. 

\section{Conclusion} \label{sec:conclusion}
Here we reported the discovery of a source of radio bursts discovered by the \realfast\ system at VLA during commensal operation on the VLA. Using \realfast\, we were also able to localize the bursts to subarcsecond precision. Follow-up observations with GBT detected more bursts but did not yield any clear periodicity. The brightness temperature and duration of the bursts suggest a compact source with a coherent emission mechanism. All the bursts at 1.4 GHz shows significant scattering, which is consistent with what is expected from the DM. However, the scattering timescale changed significantly between the first and second epochs, suggesting a dynamically evolving scattering screen. The bursts shows an average RM of 816 \rmunits.
All bursts showed high linear polarization and significant circular polarization was detected in one burst. We compared the field of this source with multi-wavelength catalogs, finding that the \realfast\ bursts appear confidently linked to the \textit{Fermi} source, and are spatially coincident with a PAN-STARRS star (however, the data provide only inconclusive confidence in their relationship). This also revealed an XMM-Newton source 1.9$''$ away from the \realfast\, position, nearby but not consistent with the burst source' position. 
Overall, it is expected that the burst source is of neutron star origin based on its clear similarities to the variety of radio-emitting pulsar phenomena.

Overall, the polarization property is consistent with most magnetars, single pulses from MSPs and young pulsars, and RRATs. The flat or gradually changing behavior of polarization angle and the radio energies are similar to the population of young pulsars and magnetars. However, the high $\gamma$-ray luminosities favor a young pulsar or a binary origin of J1818--1531. The source could be a magnetar, however this will be the first detection of a magnetar in GeV energies. Targeted X-ray follow-up is required to strengthen the magnetar origin and an optical follow-up would best explore the source as a binary, however further monitoring of pulsed radio behaviors may improve the chances at detecting a periodicity in this source and providing identification via pulsar timing.

\section{Acknowledgements}
RAT and SBS were supported in this work by NSF award \#1714897. For this work FKS acknowledges support through the NASA \textit{Fermi}/GI program cycle 12, grant 80NSSC19K1508, under which the discovery VLA observation was conducted. SBS gratefully acknowledges the support of a Sloan Fellowship. \realfast\, is supported by the NSF Advanced Technology and Instrumentation program under award 1611606. This research has made use of the NASA/IPAC Extragalactic Database (NED), which is operated by the Jet Propulsion Laboratory, California Institute of Technology, under contract with the National Aeronautics and Space Administration. Parts of this work used Ned Wright's extremely useful online cosmology calculator \citep{cosmocalc}. Computational resources were provided by the WVU Research Computing Thorny Flat HPC cluster, which is funded in part by NSF OAC-1726534. The Green Bank and National Radio Astronomy Observatories are facilities of the National Science Foundation operated under cooperative agreements by Associated Universities, Inc. We thank the telescope operators and project friends during all our VLA and GBT observations. We also thank Kaustubh Rajwade, Natasha Hurley-Walker and Matthew Kerr for useful discussion about this work.

\bibliography{fermi}{}
\bibliographystyle{aasjournal}



\end{document}